\documentclass[draftclsnofoot,onecolumn,12pt]{IEEEtran}
\usepackage{ifpdf}
\usepackage{graphicx}
\usepackage{amssymb}
\usepackage{amsmath}
\usepackage{cite}
\usepackage{epstopdf}
\usepackage{dsfont}
\usepackage[justification=centering]{caption}
\usepackage{color}
\usepackage{algorithm}
\usepackage{algorithmicx, algpseudocode}
\usepackage{etoolbox}
\usepackage{subcaption}

\makeatletter
\def\ScaleIfNeeded{%
\ifdim\Gin@nat@width>\linewidth \linewidth \else \Gin@nat@width \fi
} \makeatother

\begin{document}

\title{IRS-Empowered Wireless Communications: State-of-the-Art, Key Techniques, and Open Issues}

\author{\IEEEauthorblockN{Ming Zeng,
Ebrahim Bedeer, Xingwang Li, Quoc-Viet Pham, Octavia A. Dobre, \emph{Fellow, IEEE}, Paul Fortier, Leslie A. Rusch, \emph{Fellow, IEEE}
}
\thanks{
M. Zeng, P. Fortier, and L. A. Rusch are Universit\'e Laval, Quebec, QC, G1V 0A6, Canada (e-mail: ming.zeng, paul.fortier, leslie.rusch@gel.ulaval.ca)

E. Bedeer is with University of Saskatchewan, Saskatoon, S7N5A9, Canada, (email: e.bedeer@usask.ca)

X. Li is with Henan Polytechnic University, Jiaozuo, 454000, China (e-mail: lixingwangbupt@gmail.com)

Q. V. Pham is with Pusan National University, Busan, 46241, South Korea (e-mail: vietpq@pusan.ac.kr)

O. A. Dobre is with Memorial University, St. John's, NL, A1B 3X9, Canada (e-mail: odobre@mun.ca).

} 
}

%
%
%


\maketitle

\begin{abstract}
In this article, we overview intelligent reflecting surface (IRS)-empowered wireless communication {\color{black} systems}.
We first present the fundamentals of IRS-assisted wireless transmission.
On this basis, we explore the integration of IRS with various advanced transmission technologies, such as millimeter wave, non-orthogonal multiple access, and physical layer security. 
Following this, we discuss the effects of hardware impairments and imperfect channel-state-information on the IRS system performance. Finally, we highlight several open issues to be addressed. 


\end{abstract}
%
\IEEEpeerreviewmaketitle


\section{Introduction}
Recently, intelligent reflecting surface (IRS) has drawn great attention as a promising physical layer transmission technology for next-generation communication systems \cite{Capacity_ref_12, Capacity_ref_4, Capacity_ref_7, Capacity_ref_15}. An IRS is a planar surface equipped with massive low-cost passive reflecting
elements; each can induce a phase and/or amplitude change to the impinging signals to achieve fine-grained reflective beamforming. By judiciously deploying an IRS in the environment, an extra communication link that goes through the IRS can be built between the {\color{black}transmitter} (Tx) and receiver (Rx), and thus, better support diverse user requirements, such as extended coverage, increased data rate, minimized power consumption, and enhanced secure
transmissions \cite{Capacity_ref_12, Capacity_ref_4, Capacity_ref_7, Capacity_ref_15}. 
Not only theoretically attractive, IRS also possesses various advantages in terms of practical implementation. It is of low hardware and energy cost, and can be easily deployed on the environment objects, e.g., the facades of buildings. Moreover, IRS can operate in full-duplex mode without self-interference and noise amplification.

To fully reap the benefits provided by IRS, it is necessary to investigate the integration of IRS with other transmission technologies for next-generation communication systems,
such as millimeter wave (mmWave), non-orthogonal multiple access (NOMA), and physical layer security (PLS). 
The integration of IRS into mmWave is natural, since 
IRS can establish additional line-of-sight (LoS) links to extend the coverage of mmWave, which suffers from severe signal attenuation and poor diffraction \cite{mmWave_ref_6, mmWave_ref_10, mmWave_ref_3}. 
The application of IRS into NOMA is also promising, as IRS can be utilized to introduce desirable channel gain differences among users as well as {\color{black} to suppress} the inter-user interference; both can lead to performance improvement of NOMA systems \cite{NOMA_ref_13, NOMA_ref_16}. 
Lastly, IRS can be used to enhance the signal strength at the legitimate users while nulling the signal reception at the eavesdroppers, and thus, improve PLS of wireless systems \cite{PLS_ref_9, PLS_ref_3,  PLS_ref_8 }. 

The aim of this article is to provide a comprehensive survey on IRS-assisted wireless transmission. 
We cover both conventional IRS-assisted systems and more advanced ones where IRS is integrated with other candidate technologies, such as mmWave, NOMA and PLS. 
For all the considered scenarios, we not only present the state-of-the-art research progress, but also identify the challenges and key techniques for resource allocation. Motivated by the practical challenges of implementing IRS, 
we further discuss the effects of hardware impairments (HWIs) and imperfect channel-state-information (CSI) on the IRS system performance. Finally, several open issues are highlighted. 


\section{IRS-Assisted Wireless Transmission}
\subsection{Motivation}
Conventional network optimization in wireless communication systems has been limited to transmission control at the transceivers, with little attention has been paid to the wireless propagation environment. Indeed, the wireless propagation environment has long been perceived as an uncontrollable
and randomly behaving entity between the transceivers. Aside from being uncontrollable, the
environment usually has an adverse impact on the communication efficiency, owing to the
signal attenuation, fading, and interference introduced. As a result, the propagation environment
itself becomes a major limiting factor that hinders further performance improvement of wireless networks. Recently, there is an increasing demand for novel communication paradigms that can smartly tune the propagation environment either to increase the communication efficiency or to simplify the transceiver architecture. In this regard, IRS has received great attention owing to its ability to reconfigure the propagation environment via software controlled reflection \cite{Capacity_ref_12, Capacity_ref_4, Capacity_ref_7, Capacity_ref_15}. As shown in Fig.~1, an IRS is a planar surface consisting of many low-cost passive reflecting elements; each can induce a phase and/or amplitude change to the incident signal to achieve fine-grained reflective beamforming. When the direct link between the transceivers fails due to unfavourable channel conditions, IRS can be deployed to build a cascaded link and resumes the communication, as illustrated in Fig. 1(a). 
Even when the direct link exists, IRS can still be used to add an extra communication link between the transceivers to improve the system performance, as presented in Fig. 1(b).

\begin{figure*}
\centering
\includegraphics[width=1\linewidth]{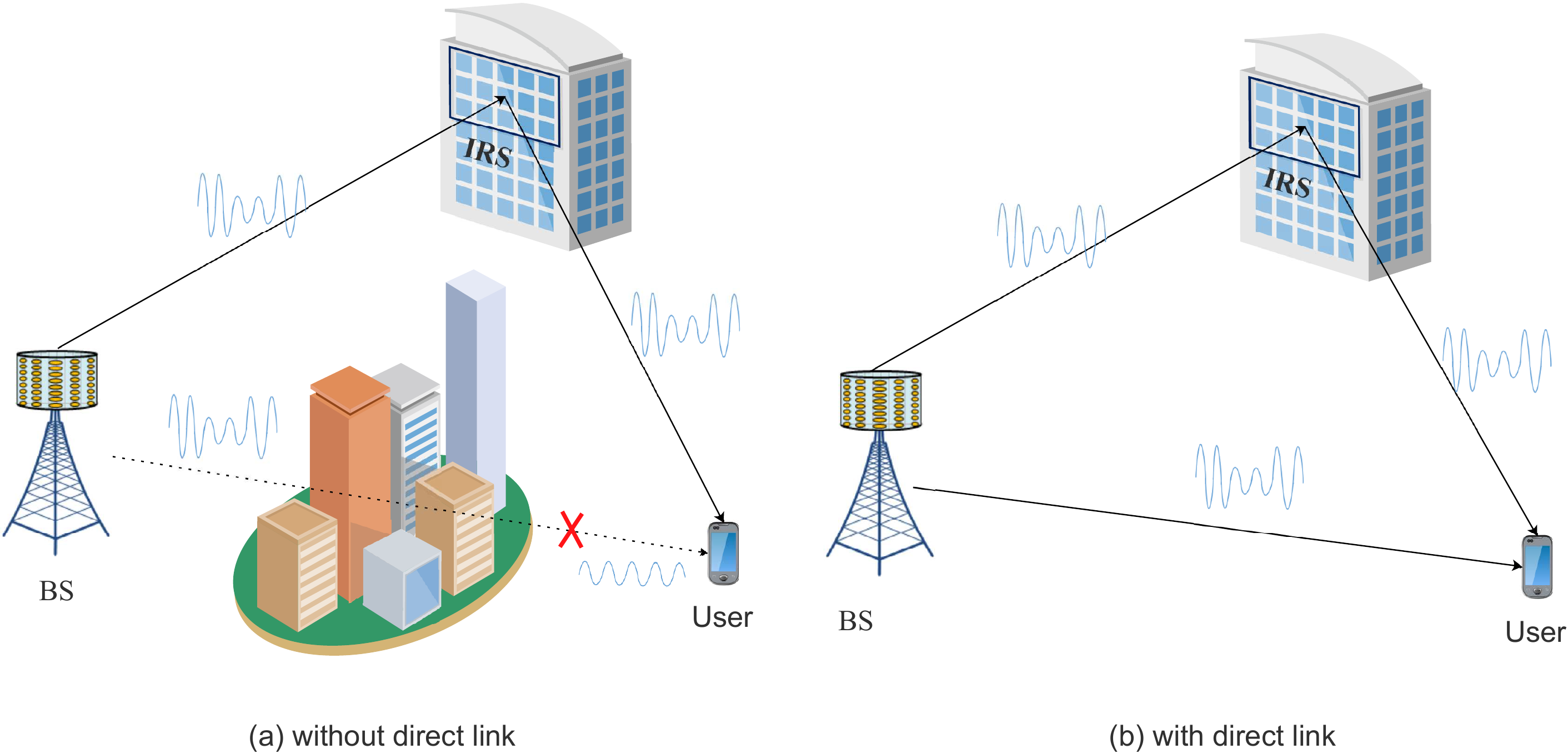}
\caption{{IRS-assisted wireless transmission: a) without direct link; b) with direct link.}}
\label{systemmodel1}
\end{figure*}

%

\subsection{State-of-the-Art and Key Techniques}
The simplest IRS-assisted communication system consists of three nodes, namely the Tx, Rx, and IRS, which is referred to as the point-to-point communication. Depending on how many antennas the transceivers are equipped with, such a system can be further classified into three categories, i.e., single-input single-output (SISO), multiple-input
single-output (MISO) and multiple-input multiple-output (MIMO) systems. 
Existing works on MISO systems have revealed that IRS can achieve squared power gain under asymptotically large number of reflecting elements \cite{Capacity_ref_12}. Such a gain shows great potential of IRS and its superiority over conventional massive MIMO systems, where only a linear gain is achieved \cite{Capacity_ref_12}. It is further shown that by judiciously controlling the IRS reflection coefficients, the IRS-assisted MIMO channel can be substantially enhanced in terms of channel power, condition number, or rank. 

In addition to point-to-point communication, IRS-assisted multi-user systems have also been studied \cite{Capacity_ref_12, Capacity_ref_4}. The performance analysis of multi-user systems is undoubtedly much more challenging than that of a point-to-point system, owing to the existence
of inter-user interference. To evaluate the fundamental capacity limits of IRS-assisted multi-user
SISO systems, \cite{Capacity_ref_4} characterizes the capacity and rate regions subject to the constraint on IRS reconfiguration times. Presented simulation results show that the capacity and rate regions can be greatly improved using IRS. 
The more general MISO scenario is studied in \cite{Capacity_ref_12}, and it is shown that employing IRS can significantly improve the coverage, energy consumption and achievable rate of wireless networks. 

The works mentioned above focus on single-cell systems. The use of IRS in multi-cell systems is however investigated in \cite{Capacity_ref_7, Capacity_ref_15}. Compared with the single-cell counterpart, new issues emerge in multi-cell systems, such as where to deploy the IRSs and how to coordinate available resources among different base stations (BSs). The authors in \cite{Capacity_ref_7} consider a large-scale
deployment of IRSs in wireless networks, and characterize the achievable spatial throughput
averaged over both channel fading and random locations of the deployed BSs/IRSs. It is unveiled
that deploying distributed IRSs can greatly boost the received signal power but only cause
marginal extra interference in the network. Note that SISO is assumed in \cite{Capacity_ref_7}. The authors in
\cite{Capacity_ref_15} study the weighted sum rate maximization problem for a MIMO multicell
system. Numerical results reveal that employing IRSs can notably enhance the cell-edge
performance.

In IRS-assisted systems, the IRS phase shifts need to be optimized in addition to the conventional
transceiver optimization. As the IRS-assisted user channels are cascaded, the variables to be optimized are often coupled, and thus, resulting in non-trivial joint resource optimization. Moreover, the IRS optimization needs to satisfy the highly non-convex constant modulus constraint, since the IRS can only reflect the incident signal without amplifying it.
Existing works often apply the block coordinate descent (BCD) method (also referred to as the
alternating optimization (AO) when there are only two types of variables) to resolve the coupling
among the optimization variables \cite{Capacity_ref_12, Capacity_ref_15}. On this basis, existing approaches developed for systems without IRS often can be borrowed for the transceiver optimization.
Meanwhile, the semidefinite relaxation (SDR) technique is
widely used for addressing the passive beamforming at the IRS. The framework combining BCD/AO with SDR is shown to be effective in handling the joint resource allocation of various IRS-assisted systems \cite{Capacity_ref_12}. Nonetheless, it still suffers from two drawbacks: i) the obtained solution can only be considered as a lower bound; and ii) the complexity may be too high, since
the high complexity SDR operation needs to be performed many times until convergence. To obtain a tight upper bound, a potential solution is to apply the successive convex approximation (SCA) technique to construct a convex approximation for the joint optimization. This however could be non-trivial due to the coupling among the variables. Nevertheless, for certain cases, by exploiting the closed form solutions existing for the transceivers optimization under given IRS phase shifts, the problem could be simplified and it becomes relatively easy to apply the SCA technique. Additionally, the following two approaches could be adopted to replace SDR and thus, lower the complexity; one is SCA while the other is the complex circle manifold (CCM) method. For SCA, the majorization-minimization (MM) algorithm appeals to be quite promising and the
key then will be to find the appropriate surrogate function \cite{Capacity_ref_15}. The usage of CCM on the other hand is motivated by the complex forms of the IRS phases, and the main challenge lies in how to design a gradient descent algorithm based on the manifold space \cite{Capacity_ref_15}. 

\section{Integration of IRS with Advanced Transmission Technologies}
To further exploit the potential of IRS, it is of interest to investigate the integration of IRS with other advanced transmission technologies, including mmWave, NOMA, and PLS. 

\subsection{IRS-assisted mmWave Communication}
\subsubsection{Motivation}
mmWave communication has drawn considerable
attention recently owing to its ability to provide ultra-wide bandwidth \cite{mmWave_ref_6, mmWave_ref_10, mmWave_ref_3}. Nonetheless, mmWave communication suffers from severe signal attenuation and poor diffraction, which significantly limits its applications in mobile cellular systems. MIMO represents an effective technology to enhance the mmWave signal strength owing to the high beamforming gain. However, the property of poor diffraction still makes mmWave vulnerable to blocking {\color{black} by} obstacles that break the LoS links. To address this, IRS can be deployed to create additional LoS links, and thus, extend the coverage of mmWave systems \cite{mmWave_ref_6, mmWave_ref_10, mmWave_ref_3}. 

\subsubsection{State-of-the-Art and Key Techniques}
\begin{figure}
\centering
\includegraphics[width=1\linewidth]{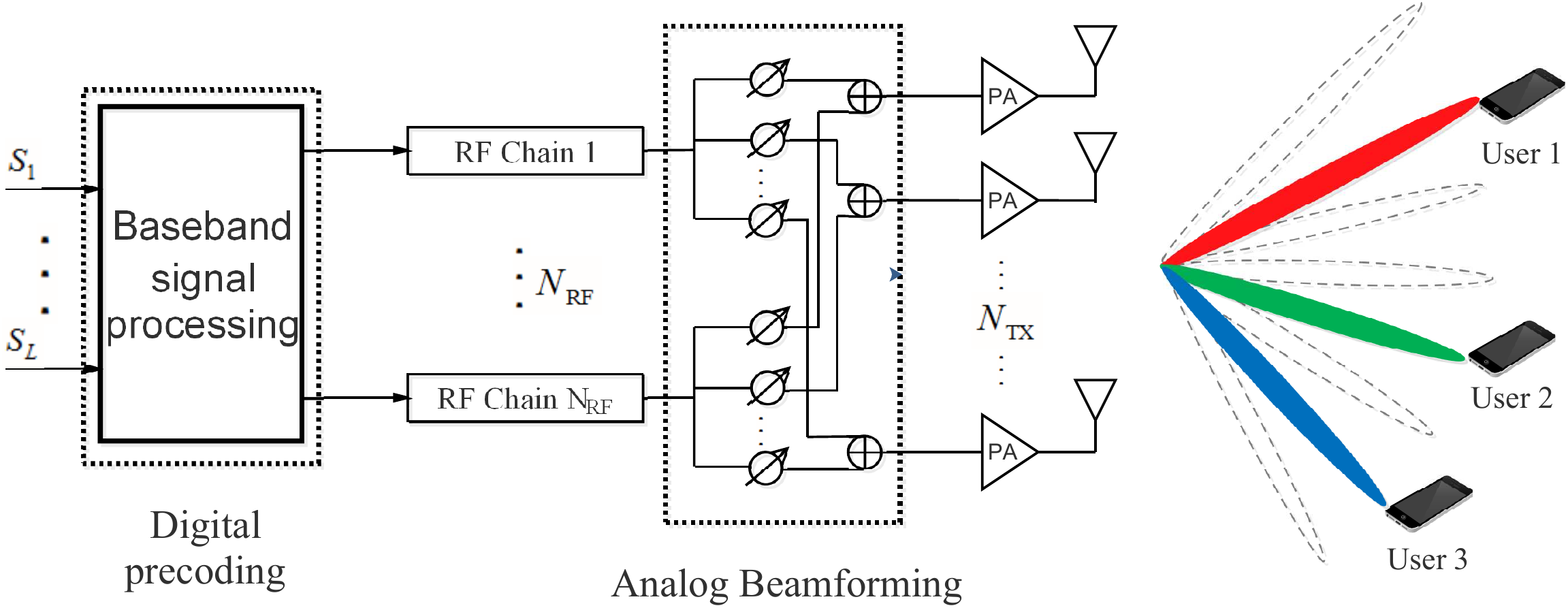}
\caption{{Structure of hybrid analog/digital beamforming.}}
\label{hybrid}
\end{figure}

\begin{figure}
\centering
\includegraphics[width=1\linewidth]{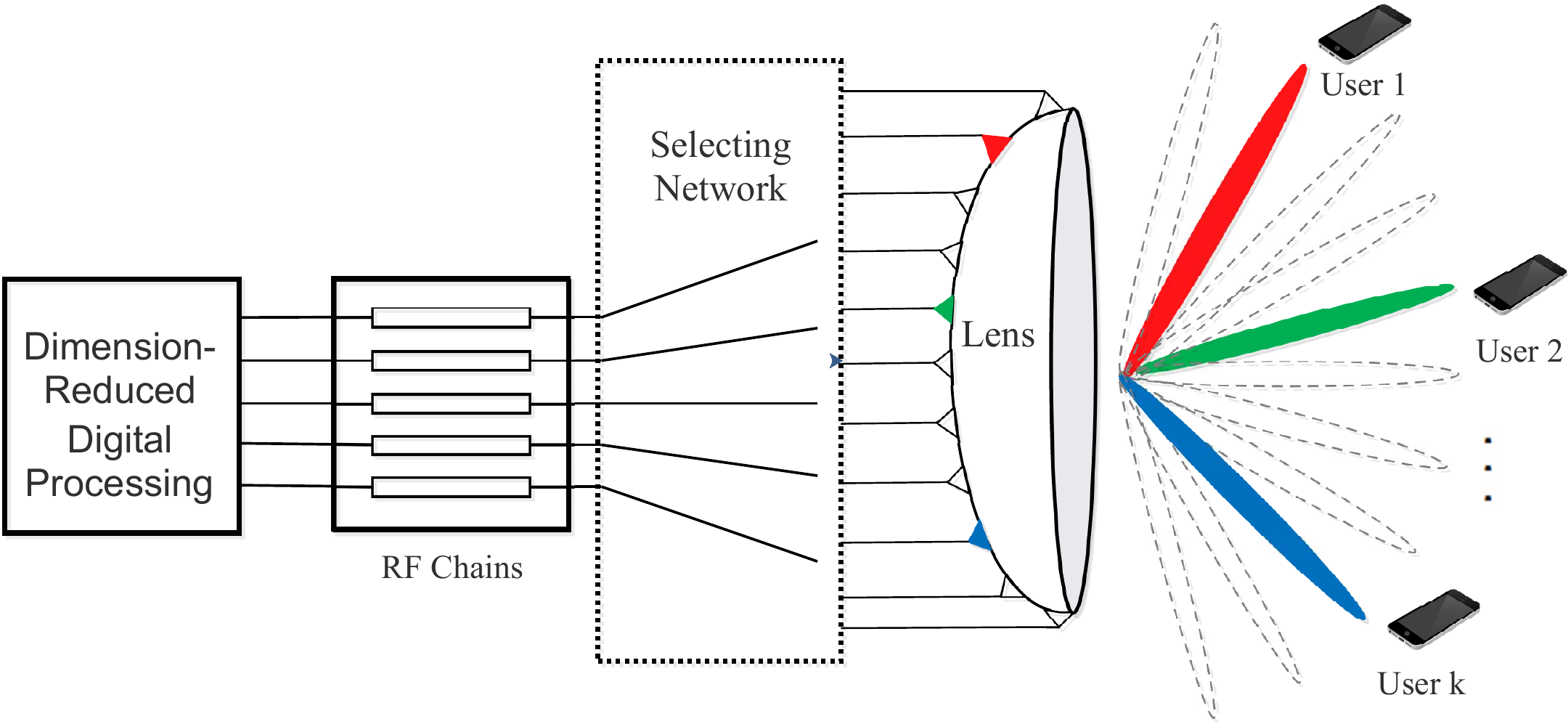}
\caption{{Structure of lens antenna array.}}
\label{lens}
\end{figure}

Earlier works on IRS-assisted mmWave MIMO focus on full digital beamforming at the BS, and have shown that deploying IRS can alleviate the blockage effect and enhance the performance of mmWave systems in coverage and throughput \cite{mmWave_ref_6}.  
To lower the number of RF chains, one can either adopt the hybrid analog/digital beamforming structure (shown in Fig.~\ref{hybrid} \cite{mmWave_ref_10}) or the lens antenna array (illustrated in Fig.~\ref{lens} \cite{mmWave_ref_3}). The former consists of two parts, namely the baseband digital beamforming under limited number of RF chains and the RF band analog beamforming via a network of phase shifters. Likewise, the latter also comprises of two main components, namely the electromagnetic (EM) lens and the matching antenna array with elements located in the focal region of the lens. The EM lenses provide controllable phase shifting to obtain angle-dependent energy focusing property.
It is shown that using IRS can enhance the performance for both hybrid beamforming and lens antenna array based mmWave systems \cite{mmWave_ref_10, mmWave_ref_3}. 

The BCD framework for microwave can be applied to mmWave as well. However, the channel model should be updated using the appropriate mmWave channel models, e.g., the geometric channel model \cite{mmWave_ref_6, mmWave_ref_10}. Considering the poor diffraction and penetration abilities of mmWave, the direct links between the users and BS under mmWave are often assumed blocked \cite{mmWave_ref_10, mmWave_ref_3}. Furthermore, the Tx-IRS and IRS-Rx channels can be approximated as a rank-one matrix/vector, as they are LoS dominated. This rank-one structure can be exploited for further simplifying the passive beamforming design at the IRS \cite{mmWave_ref_6} and the active beamforming at the BS \cite{mmWave_ref_10}. Additionally, the analog beamforming under hybrid beamforming in general can be handled using SDR as the phase shift optimization at the IRS, since both are limited by the same constant-modulus constraint. To lower the complexity, beam search based on pre-defined codebooks can also be adopted. In terms of lens antenna array, the key lies in how to perform an appropriate antenna/beam selection to significantly lower the RF chain cost, without sacrificing the system performance too much \cite{mmWave_ref_3}. Interference-aware beam selection could be of interest.

\subsection{IRS-assisted NOMA Transmission}
\subsubsection{Motivation}
\begin{figure}
\centering
\includegraphics[width=0.8\linewidth]{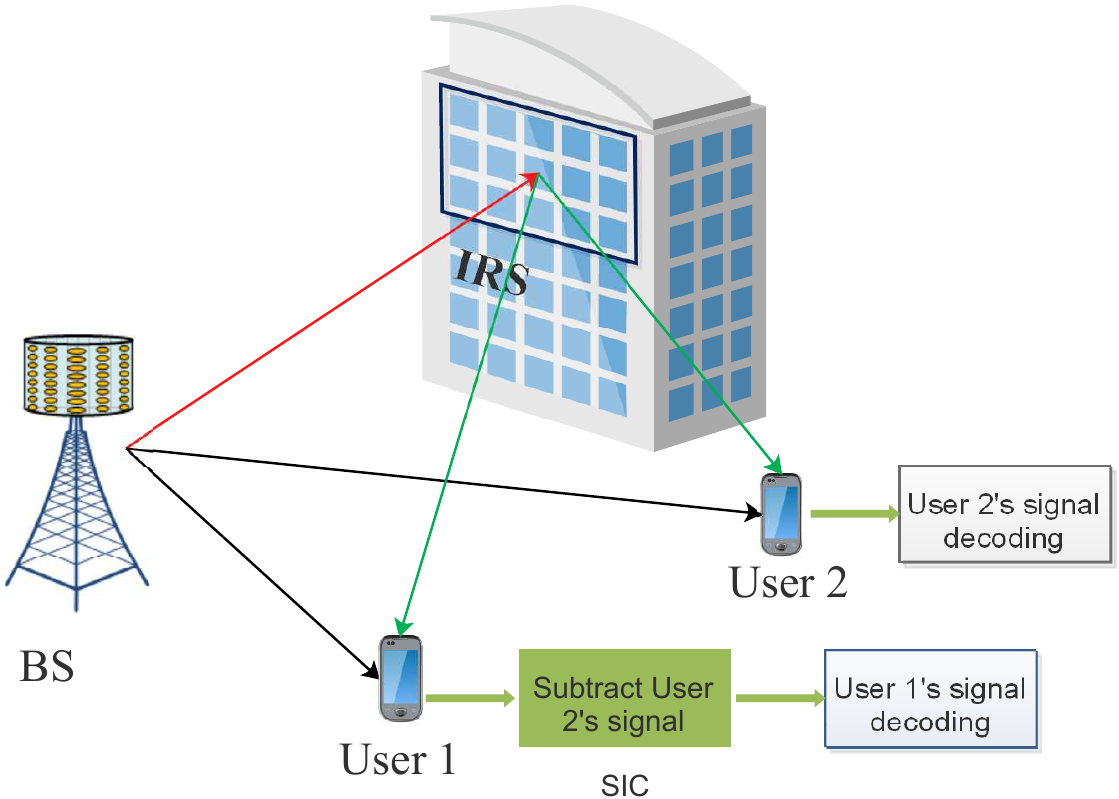}
\caption{{Illustration of IRS-assisted NOMA transmission.}}
\label{NOMA}
\end{figure}

NOMA is envisioned as a promising radio access technique for next-generation communication
systems \cite{NOMA_ref_13, NOMA_ref_16}. By enabling multiple users to access the same time/frequency resources, NOMA can achieve higher spectral-efficiency and energy efficiency and better support massive connectivity when compared to orthogonal multiple access (OMA). Nevertheless, to obtain a decent performance gain of NOMA over OMA, the users are required to have a large channel gain
disparity. Such a requirement may be violated in conventional NOMA systems, since the
user channels are determined by the highly stochastic propagation environments.
To overcome this, IRS can be utilized to introduce desirable channel gain differences among
the users via constructively or destructively adding the user signals (shown in Fig.~\ref{NOMA}). Meanwhile, IRS can also
be used to suppress the inter-user interference, and thus, lead to improved throughput or fairness
of NOMA systems.

\subsubsection{State-of-the-Art and Key Techniques}
A body of research works has emerged very recently which investigate the design of IRS-assisted NOMA systems \cite{NOMA_ref_13, NOMA_ref_16}.
The authors in \cite{NOMA_ref_13} investigate the sum rate maximization for an IRS-assisted multi-user SISO-NOMA system, requiring to jointly optimize the channel assignment, decoding order of NOMA users, power allocation, and reflection coefficients. A three-step resource allocation algorithm is proposed and
presented numerical results demonstrate the superiority of IRS-assisted NOMA over conventional NOMA without
IRS and IRS-assisted OMA in terms of system throughput. 
To further enhance the system performance,  \cite{NOMA_ref_16} aims to maximize the minimum rate of all users for an IRS-assisted MIMO-NOMA system, by jointly optimizing the transmit beamforming at the BS and the phase shifts at the IRS. An efficient algorithm based on the framework of BCD with SDR is proposed to address the formulated non-convex problem. 
It is shown that the IRS-assisted MIMO-NOMA system can greatly boost the rate performance, when compared with conventional NOMA without IRS and OMA with/without IRS. 

In IRS-assisted NOMA systems, resource allocation becomes more complicated due to the extra need for optimizing the decoding order \cite{NOMA_ref_16}. Moreover, 
the optimization of the decoding order
is coupled with that of the IRS phase shifts, since the optimal decoding order cannot be determined without knowing the IRS phase shifts while the IRS phase shifts cannot be properly configured without fixing the decoding order. 
To decompose the coupling among them, a viable solution could be to iteratively update the decoding order and IRS phase shifts, by fixing the other. However, the complexity may become prohibited when the number of iterations required for convergence is large. 
An alternative is to exhaustively search all decoding orders and on this basis, optimize the IRS phase shifts. Likewise, the resulting complexity of exhaustive search may be too high, especially when the number of users is large.
To lower the complexity, one can greedily set the decoding order by fixing the IRS phase matrix to certain values, such as all zeros or ones. When the IRS phase matrix is set to all zeros, it means only the direct link is considered. Clearly, this may be highly suboptimal due to the neglect of the effect of the cascaded Tx-IRS-Rx link. In contrast, when the IRS phase matrix is set to all ones, both the direct link and the cascaded Tx-IRS-Rx link are considered. Nevertheless, the resulting decoding order may still be quite different from the optimal one for the optimized IRS phase matrix. To address this, the authors in \cite{NOMA_ref_16} propose a combined-channel strength-based user
ordering scheme, where users are ordered based on their maximally achievable combined channel strengths via optimizing the IRS phase shifts. Numerical results show that the proposed user ordering scheme achieves near-optimal performance with much lower complexity. 
Except for user ordering, extra constraints should be imposed on users' achievable rates in IRS-assisted NOMA systems, to ensure the success of successive interference cancellation (SIC). 
That is, each user's achievable rate cannot exceed the minimum rates decodable at all users that need to decode its signal. 
Such extra constraints may make it more challenging to identify the initial feasible IRS phase shifts required for the BCD based optimization framework.  

\subsection{IRS-assisted PLS Systems}

\subsubsection{Motivation}
Owing to the broadcasting nature of wireless transmission media, wireless systems are vulnerable to 
impersonation attacks and eavesdropping. 
Encryption techniques represent an effective way to ensure communication confidentiality; however, they
may not be suitable for some Internet-of-things applications that cannot afford their complexity
and/or have stringent delay requirements.
Through exploiting the randomness nature of wireless
propagation channels, PLS can help secure wireless communication confidentiality without
consuming much of the resources.
To ensure a non-negative secrecy rate, it is however often required that the legitimate users experience better
channel conditions than the eavesdroppers. Clearly, such a requirement does not always hold in
conventional PLS systems. A simple workaround is to deploy IRS in PLS systems to reconfigure
the channels for the legitimate users and eavesdroppers. In particular, IRS can be used to enhance
the signal strength at the legitimate users while nulling the signal reception at the eavesdroppers, thereby enhancing the secrecy transmission rates.

\subsubsection{State-of-the-Art and Key Techniques}
\begin{figure}
\centering
\includegraphics[width=0.8\linewidth]{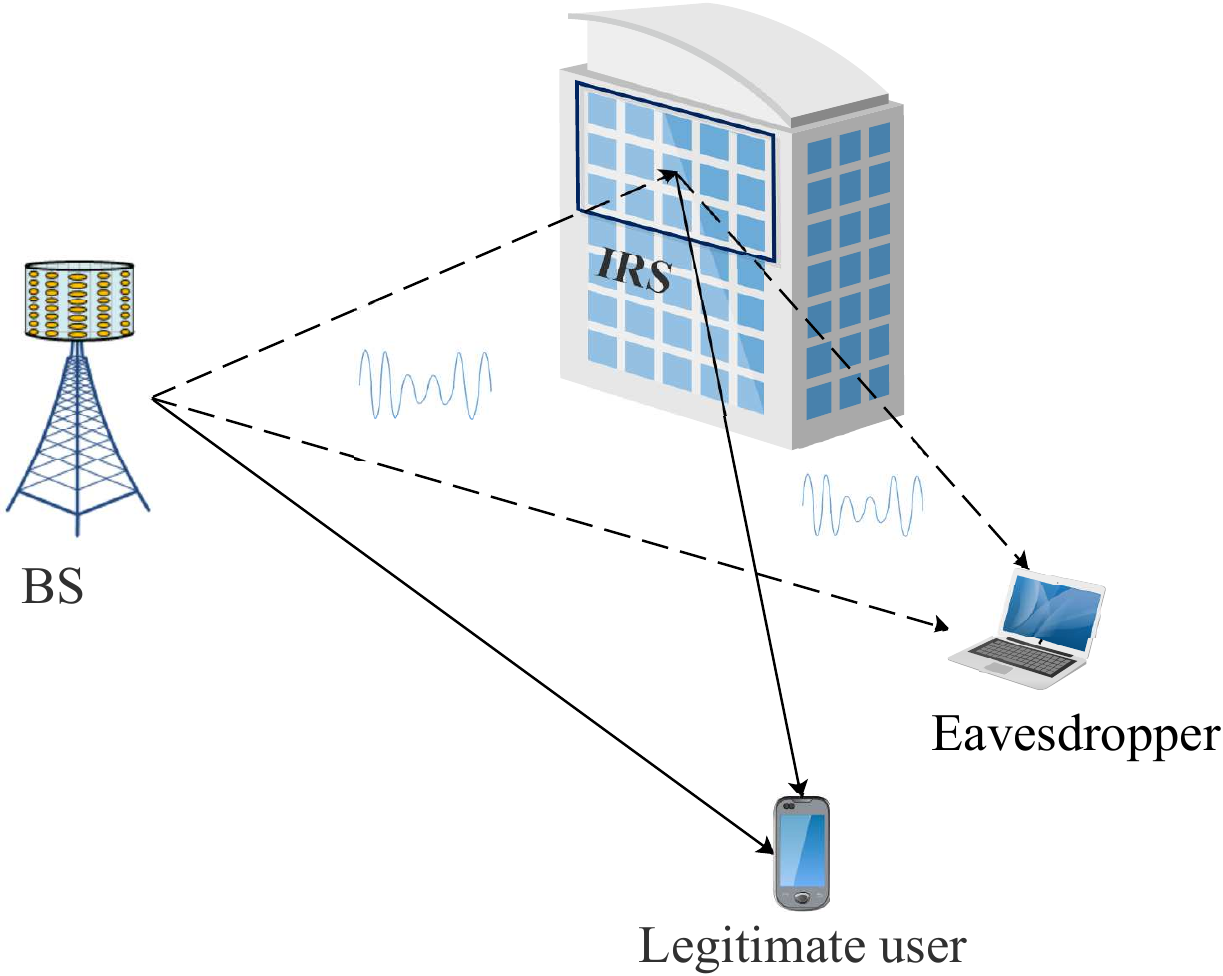}
\caption{Illustration of IRS-assisted PLS systems.}
\label{PLS}
\end{figure}

The research on IRS-assisted PLS systems is still at an earlier stage, and existing works mainly focus on simple scenarios with one legitimate user and one eavesdropper (shown in Fig.~\ref{PLS}), e.g., \cite{PLS_ref_9, PLS_ref_3, PLS_ref_8}. Both \cite{PLS_ref_9} and \cite{PLS_ref_3} consider the case where the BS is equipped with multiple antennas, while the legitimate user and eavesdropper are with single antenna. 
\cite{PLS_ref_9} aims to maximize the secrecy transmission rate under the maximum transmit power constraint, whereas \cite{PLS_ref_3} studies the transmit power minimization subject to secrecy rate constraint at the legitimate user. Presented simulation results show that the proposed schemes with IRS outperform their counterparts without IRS in terms of secrecy rate and transmit power. 
In addition, the authors in \cite{PLS_ref_8} analyse the secrecy outage probability of an IRS-assisted PLS system, where all network nodes are equipped with single antenna. 
Numerical results illustrate that
deploying the IRS can lower the secrecy outage probability as well.

Introducing PLS into IRS-assisted systems often leads to the original non-convex objective function being more complicated, and thus harder to handle. As in conventional IRS systems, the BCD method can be used to decompose the coupling among the optimization variables, thereby making the problem more tractable. On this basis, there exist several ways to optimize the transmit beamforming. First, a closed-form solution may be derived for the optimal transmit beamformer, e.g., \cite{PLS_ref_9, PLS_ref_3}. Second, the SCA technique, e.g., difference-of-convex (DC) programming, could be used, considering the expression of the secrecy rate. Last, fractional programming may also be employed, by removing the $\log(\cdot)$ operation in the objective function. For the IRS phase shifts optimization, semi-closed form solutions may exist in certain cases, e.g., \cite{PLS_ref_9}. Besides, SDR has been widely used to obtain a high quality solution, e.g. \cite{PLS_ref_3}. Last, the SCA technique, e.g., the MM algorithm can also be adopted. 

\begin{figure} 
\centering
\includegraphics[width=1\textwidth]{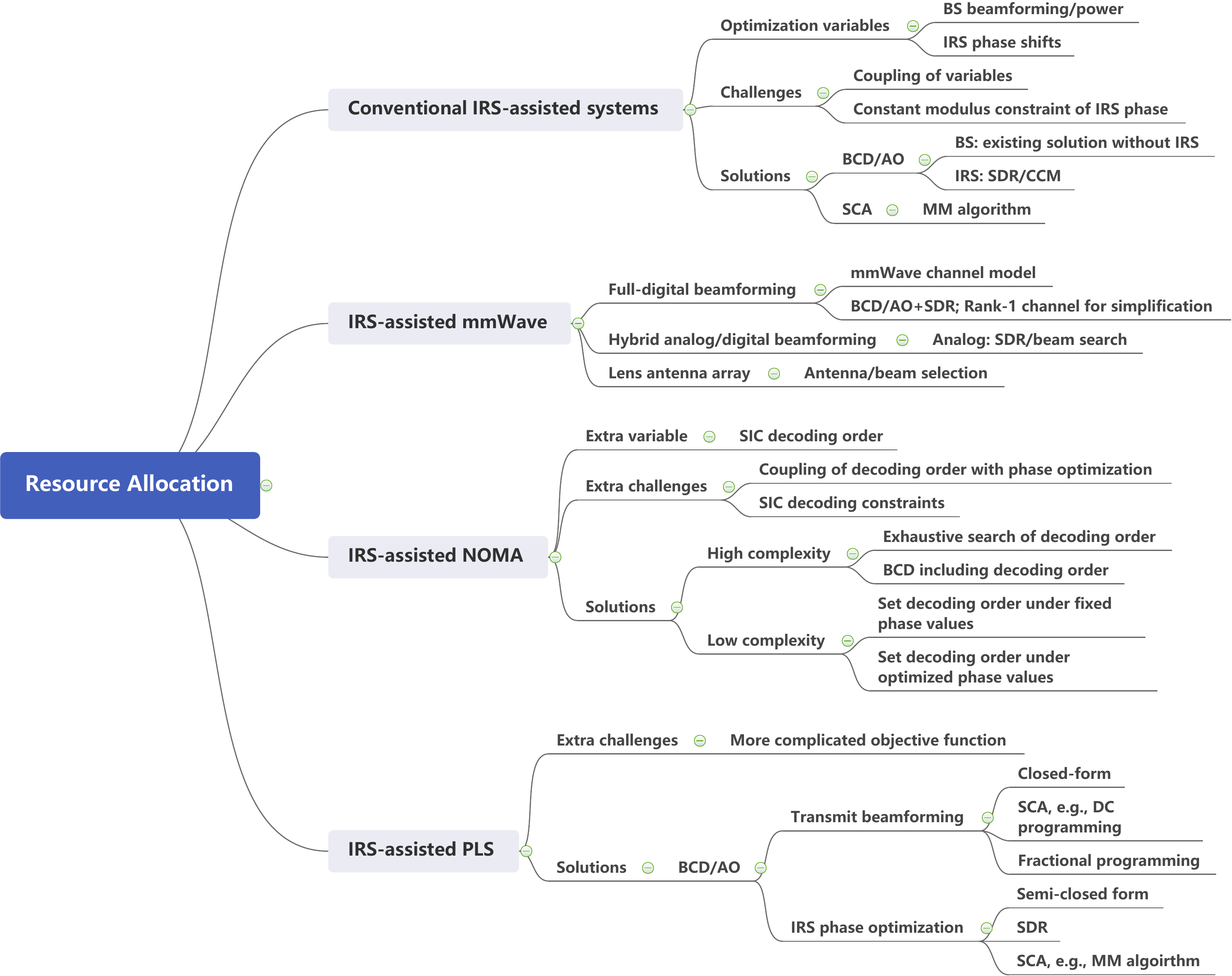}
\caption{\small{Summary of research allocation for conventional and advanced IRS-assisted systems.}}
\label{RA}
\end{figure}

A summary of resource allocation for conventional and advanced IRS-assisted systems is given in Fig.~\ref{RA}.

\section{Non-Ideal Transmission of IRS-Assisted Communication Systems}
IRS-assisted communication systems, like any other communication systems, suffer from non-ideal transmission conditions (mainly result from HWIs and/or imperfect CSI) that can deteriorate their performance if not properly taken into consideration. HWIs in IRS-assisted communications systems can result from the finite-resolution of the phase shifters at the IRS reflecting elements and/or  the RF front-end mismatches at the transceivers. In the rest of this section, we will discuss HWIs and imperfect CSI and their effect on the performance of IRS-assisted systems.

\subsection{Hardware Impairments}
\subsubsection{Finite-Resolution Phase Shifters}
IRS reflecting elements need to adjust their phase shifts in real time to compensate for the time-varying nature of the wireless channel. Such phase shift adjustments of IRS reflecting elements can be achieved via using positive intrinsic-negative (PIN) diodes, micro-electromechanical system based switches, or field-effect transistors \cite{nayeri2018reflectarray}. A large number of studies on IRS consider that the phase shifts of the IRS reflecting elements can change continuously, which is hard to achieve in practice. For instance, the phase shift levels of the PIN diode is typically adjusted to two levels (0 and $\pi$ radians) by changing the applied biasing voltage between two levels. To achieve $M$ different phase shift levels, $\log_2 M$ PIN diodes are required for each reflecting elements. Alternatively, a single varactor diode can be used to achieve more than two phase shifts; however, it requires a high number of biasing voltages which will increase the complexity of the IRS controller. 
Manufacturing IRS reflecting elements to support a high number of phase shifts will increase its cost, and hence, will not be a scalable solution given that IRS typically has a very high number of reflecting elements. The effect of the finite-resolution phase shifters of the IRS reflecting elements is investigated in a number of recent works. It is shown that for an asymptotically large number of IRS reflecting elements, a 1-bit phase shifter approaches the same squared power gain when compared to the ideal continuous phase shifters. However, as the number of IRS reflecting elements decreases, the power loss increases and it depends on the number of the available phase shift levels.

\subsubsection{RF Chain Impairments}
RF font-end impairments in IRS-assisted communication systems include in-phase and quadrature imbalance (IQI) at the transceiver, phase noise of IRS, and other transmission non-linearities. A general model to capture such RF impairments in IRS-assisted communication systems is the extended error vector magnitude (EEVM) proposed in \cite{schenk2008rf}. In the EEVM model, the RF chain impairments are effectively modeled as multiplicative and additive random variables terms, with a certain mean and variance. This model has been applied to both single- and multi-antenna transmitters in IRS-assisted communication systems. It is shown that the aggregate RF front-end impairments at both Tx and Rx can be accurately modeled as a zero-mean complex Gaussian process with a variance that depends on the transmit power, the effective channel gain of the Tx-IRS-Rx, and the square of the error vector magnitude at both the Tx and Rx. Analysis of IRS-assisted communication systems employing the EEVM model shows that at high signal-noise-ratio, the performance (in terms of spectral efficiency or outage probability) is independent of the number of IRS reflecting elements, and is mainly constrained by the level of the RF chain impairments at the Tx when compared to the phase distortion at the IRS elements. Such analysis reveals that modest IRS elements with low-resolution phase shifters can be used as affordable deployment solutions for IRS without significantly degrading the performance.

\subsection{Imperfect CSI}

Estimating the indirect channel (between the Tx and IRS) and the reflection channel (between the IRS and Rx) is challenging given the fact that IRS consists of passive reflecting elements with limited signal processing capabilities. The indirect channel estimation problem can be partially solved with the knowledge of the angle of arrival given that the IRS elements are mounted on buildings, and hence, considered of fixed location. The reflection channel estimation is challenging given the expected end users' mobility, and errors in the estimation of the reflection channel can significantly deteriorates the performance of IRS-assisted communication systems if their effect is not properly considered in the IRS system design. In general, imperfect CSI estimation can be addressed through a worst case/robust design, considering knowledge of the channel statistics rather than the instantaneous channel coefficients, or allowing a controlled outage in the IRS system performance. In a robust design, the transmit power will be typically increased, when compared to its counterpart assuming perfect CSI, to compensate for the estimation inaccuracy. 
If the statistics of the CSI is known, it is possible to improve the performance of IRS systems (e.g., achievable rate) on average rather than instantaneously. However, recent studies in the literature reported that the the performance of IRS systems under statistical CSI knowledge deteriorates with increasing the number of IRS elements serving a particular end user \cite{HW_ref_10}.

\section{Open Issues}
In this section we highlight several open issues that are worthy of investigation.

\subsection{Unmanned Aerial Vehicles-integrated IRS systems}
Unmanned aerial vehicles (UAVs) have shown several benefits, as relays or flying BSs, to improve the performance of communication systems, and hence, they are currently being considered as a key enabler of next-generation wireless systems. UAVs typically have a strong LoS and favorable propagation conditions to terrestrial BSs which will lead radio waves from terrestrial BSs to interfere with UAVs in adjacent cells. IRS is expected to be a promising candidate solution to mitigate the inter-cell interference problem of future UAVs networks. This is because IRS can efficiently control the direction of travel of radio waves, through joint beamforming with terrestrial BSs, to reduce the power leaked to UAVs in adjacent cells. 

\subsection{Machine Learning-empowered IRS Systems}
Due to the coupling of optimization variables and non-convex nature of the underlying problems, joint resource allocation in IRS systems is challenging to solve, and often sophisticated solutions with high complexity are required to obtain near-optimal performance. 
However, the time-varying and highly dynamic nature of wireless networks requires the proposed solutions to be of low complexity and execute easily. Such a dilemma is non-trivial to overcome using conventional optimization based methods. A promising way to tackle this is to employ machine learning techniques, which have been shown as an effective tool to obtain near-optimal solutions for non-convex and sophisticated optimization problems under highly dynamic wireless environments. Machine learning also holds the potential to learn the channel indirectly from the data during training, without the need for explicit CSI.
Accordingly, machine learning-empowered IRS systems are of practical interest.

\subsection{Sensing and Localization}
Next-generation wireless communication systems will operate at higher frequencies (mmWave or Terahertz band) to support applications that require sensing of the surrounding environment and accurate localization. Such high frequencies have a limited number of propagation paths (mainly due to large penetration losses, high values of path loss, and low scattering) which may reduce the accuracy of sensing and localization. 
Hence, IRS as a controlled and dynamic scattering is considered a promising solution to the sensing and localization problem for next-generation wireless communication systems. One of the main research challenges is that for such high frequencies and large size IRS, users are no longer in the far-field and conventional sensing and localization models are no longer valid.  That said, sensing and localization models in the near-field that exploit the information in the wavefront curvature need to be developed.

\section{Conclusion}
In this article, we surveyed IRS-empowered wireless networks. We first showed that judiciously deploying IRS can substantially improve the spectral efficiency, energy efficiency and coverage of wireless networks. On this basis, we validated that IRS can be further used to enhance the performance of mmWave, NOMA, and PLS systems. However, the promised gains of IRS are often obtained under ideal assumptions on channel estimation and hardware configuration. 
Motivated by this, we further discussed the effects of HWIs and imperfect CSI on the performance of IRS. Lastly, we identified three open issues for future research. 

%
\bibliographystyle{IEEEtran}
\bibliography{IEEEabrv,conf_short,jour_short,mybibfile}

\begin{thebibliography}{10}
\providecommand{\url}[1]{#1}
\csname url@samestyle\endcsname
\providecommand{\newblock}{\relax}
\providecommand{\bibinfo}[2]{#2}
\providecommand{\BIBentrySTDinterwordspacing}{\spaceskip=0pt\relax}
\providecommand{\BIBentryALTinterwordstretchfactor}{4}
\providecommand{\BIBentryALTinterwordspacing}{\spaceskip=\fontdimen2\font plus
\BIBentryALTinterwordstretchfactor\fontdimen3\font minus
  \fontdimen4\font\relax}
\providecommand{\BIBforeignlanguage}[2]{{%
\expandafter\ifx\csname l@#1\endcsname\relax
\typeout{** WARNING: IEEEtran.bst: No hyphenation pattern has been}%
\typeout{** loaded for the language `#1'. Using the pattern for}%
\typeout{** the default language instead.}%
\else
\language=\csname l@#1\endcsname
\fi
#2}}
\providecommand{\BIBdecl}{\relax}
\BIBdecl

\bibitem{Capacity_ref_12}
Q.~{Wu} and R.~{Zhang}, ``Intelligent reflecting surface enhanced wireless
  network via joint active and passive beamforming,'' \emph{{IEEE} Trans.
  Wireless Commun.}, vol.~18, no.~11, pp. 5394--5409, Nov. 2019.

\bibitem{Capacity_ref_4}
X.~Ma \emph{et~al.}, ``Capacity and optimal resource allocation for
  {IRS}-assisted multi-user communication systems,'' \emph{\emph{[Online].
  Available: arXiv:2001.03913}}, May 2020.

\bibitem{Capacity_ref_7}
J.~Lyu and R.~Zhang, ``Hybrid active/passive wireless network aided by
  intelligent reflecting surface: System modeling and performance analysis,''
  \emph{\emph{[Online]. Available: arXiv:2004.13318}}, Sep. 2020.

\bibitem{Capacity_ref_15}
C.~{Pan} \emph{et~al.}, ``Multicell {MIMO} communications relying on
  intelligent reflecting surfaces,'' \emph{{IEEE} Trans. Wireless Commun.},
  vol.~19, no.~8, pp. 5218--5233, Aug. 2020.

\bibitem{mmWave_ref_6}
P.~Wang \emph{et~al.}, ``Intelligent reflecting surface-assisted millimeter
  wave communications: Joint active and passive precoding design,''
  \emph{{IEEE} Trans. Veh. Technol.}, pp. 1--1, Oct. 2020.

\bibitem{mmWave_ref_10}
J.~{Qiao} and M.~{Alouini}, ``Secure transmission for intelligent reflecting
  surface-assisted mmwave and terahertz systems,'' \emph{IEEE Wireless Commun.
  Lett.}, vol.~9, no.~10, pp. 1743--1747, Jun. 2020.

\bibitem{mmWave_ref_3}
Y.~Wang \emph{et~al.}, ``Energy efficiency optimization in {IRS}-enhanced
  mmwave systems with lens antenna array,'' in \emph{Proc. IEEE GLOBECOM}, Dec.
  2020, pp. 1--6.

\bibitem{NOMA_ref_13}
J.~{Zuo} \emph{et~al.}, ``Resource allocation in intelligent reflecting surface
  assisted {NOMA} systems,'' \emph{{IEEE} Trans. Commun.}, pp. 1--1, Aug. 2020.

\bibitem{NOMA_ref_16}
G.~Yang, X.~Xu, and Y.-C. Liang, ``Intelligent reflecting surface assisted
  non-orthogonal multiple access,'' \emph{\emph{[Online]. Available:
  arXiv:1907.03133}}, Dec. 2019.

\bibitem{PLS_ref_9}
H.~{Shen}, W.~{Xu}, S.~{Gong}, Z.~{He}, and C.~{Zhao}, ``Secrecy rate
  maximization for intelligent reflecting surface assisted multi-antenna
  communications,'' \emph{IEEE Commun. Lett.}, vol.~23, no.~9, pp. 1488--1492,
  Sep. 2019.

\bibitem{PLS_ref_3}
Z.~{Chu}, W.~{Hao}, P.~{Xiao}, and J.~{Shi}, ``Intelligent reflecting surface
  aided multi-antenna secure transmission,'' \emph{IEEE Wireless Commun.
  Lett.}, vol.~9, no.~1, pp. 108--112, Jan. 2020.

\bibitem{PLS_ref_8}
L.~{Yang} \emph{et~al.}, ``Secrecy performance analysis of {RIS}-aided wireless
  communication systems,'' \emph{{IEEE} Trans. Veh. Technol.}, vol.~69, no.~10,
  pp. 12\,296--12\,300, Oct. 2020.

\bibitem{nayeri2018reflectarray}
P.~Nayeri, F.~Yang, and A.~Z. Elsherbeni, \emph{Reflectarray Antennas: Theory,
  Designs and Applications}.\hskip 1em plus 0.5em minus 0.4em\relax Wiley
  Online Library, 2018.

\bibitem{schenk2008rf}
T.~Schenk, \emph{RF imperfections in high-rate wireless systems: impact and
  digital compensation}.\hskip 1em plus 0.5em minus 0.4em\relax Springer
  Science \& Business Media, 2008.

\bibitem{HW_ref_10}
H.~Guo \emph{et~al.}, ``Model-free optimization for reconfigurable intelligent
  surface with statistical {CSI},'' \emph{\emph{[Online]. Available:
  arXiv:1912.10913}}, Dec. 2019.

\end{thebibliography}

\end{document}